\documentclass[preprint,amsmath,amssymb,aps]{revtex4-1}
\usepackage[dvipdfm]{graphicx}
\usepackage{comment,color,braket,epsfig,hyperref,here,natbib,slashed,subfigure,mathrsfs}
\graphicspath{{./Figures/}}
\def\nn{\nonumber}

\def\beq{\begin{equation}}
\def\eeq{\end{equation}}
\def\bea{\begin{eqnarray}}
\def\eea{\end{eqnarray}}
\begin{document}
\title{Is the right-handed current contribution to $\bar{B}\to X_ul\nu$ decays corrected by the nontrivial vacuum in QCD?}
\author{Hiroyuki Umeeda}
\email{umeeda@gate.sinica.edu.tw}
\affiliation{Institute of Physics, Academia Sinica, Taipei 11529, Taiwan, Republic of China}
\date{\today}
\begin{abstract}
We study violation of quark-hadron duality for $\bar{B}\to X_ul\nu$ and $D\to X_dl\nu$ decays in the presence of the right-handed current operator. For this case, we show that duality violation has some aspects different from the standard model due to the existence of an instanton. In particular, the fermionic zero mode of an intermediate light quark can give a nonvanishing contribution owing to its chirality structure. A duality-violating component that arises from a single instanton is then analytically extracted as a finite-distance singularity, leading to an oscillatory correction to the observable in Minkowski space. Finally, we show that the size of duality violation for the lepton energy distribution in $\bar{B}$ decays can be comparable to the chirally suppressed perturbative contribution although the absolute size depends on the detail of the QCD vacuum.
\end{abstract}
\maketitle
\section{Introduction}
Semileptonic decays of the $\bar{B}$ meson play a central role in determining Cabibbo-Kobayashi-Maskawa (CKM) matrix \cite{Cabibbo:1963yz} elements, $|V_{qb}|~(q=c, u)$. Experimental measurements for these quantities are implemented in two distinct processes: inclusive and exclusive decays. For the inclusive processes, the theoretical analysis is basically formulated by the operator product expansion (OPE) \cite{Wilson:1969zs}, where the observable is expanded by an inverse power of heavy quark. As for the exclusive case, it can be theoretically calculated by estimating the hadronic matrix element in $\bar{B}$ decays into specific finial states. Particularly noteworthy is that the mentioned CKM matrix elements determined by these two distinct measurements are in disagreement with one another \cite{HFLAV:2022pwe}, constituting one of the long-standing issues in heavy flavor physics. See Refs.~\cite{Crivellin:2009sd, Crivellin:2014zpa} for discussions beyond the standard model in regards to the CKM puzzles.
\par
In the inclusive analysis, it should be cautioned that quark-hadron duality \cite{Bloom:1971ye, Poggio:1975af} is tacitly assumed. Uncertainty that arises from duality violation is hard to quantify, as it parametrizes our ignorance in some nonperturbative aspects of quantum chromodynamics (QCD) \cite{Shifman:2000jv}. Because of this notorious complexity, duality violation has been modeled by certain dynamical mechanisms, including (a) the resonance-based approach and (b) the instanton-based approach. In addition, (c) the method based on lattice QCD has been proposed recently. For (a), the linear Regge trajectory and the large-$N_c$ limit are taken as starting points of discussion. In particular, a solvable aspect of two-dimensional QCD known as the 't Hooft model \cite{tHooft:1974pnl} provides a useful testing ground of quark-hadron duality. The resonance-based approach has been applied for heavy quark physics in previous works~\cite{Shifman:1994yf, Zhitnitsky:1995qa, Colangelo:1997ni, Grinstein:1997xk,Bigi:1998kc, Bigi:1999fi, Burkardt:2000ez, Grinstein:2001zq, Mondejar:2006ct, Umeeda:2021llf}.
\par
As for (b), it is noted that the short-distance expansion at the vicinity of $x^2=0$ for the Green function is incapable of reproducing the correlator in the momentum space, up to singularities of the quark propagator that reside at finite distance $(x^2\neq 0)$. Hence, those singularities give rise to an error in the OPE analysis, as shown in a one-dimensional mathematical example \cite{Chibisov:1996wf}. By taking account of possible sources of the finite-distance singularity, one can extract a correction to the OPE that is exponentially suppressed (oscillatory) in the Euclidean (Minkowski) domain. In this sense, an instanton \cite{Belavin:1975fg} is considered a probe of the duality-violating components caused by the nonperturbative effect. In the previous works, the contribution of finite-distance singularity has been discussed in Refs.~\cite{Dubovikov:1981bf, Balitsky:1992vs} and formally investigated for quark-hadron duality in Ref.~\cite{Chibisov:1996wf}.
\par
For the case of (c), it is based on the {\it ab initio} calculation of hadronic correlators in lattice QCD for inclusive $B$ decays \cite{Gambino:2020crt}. In this formalism, (partially) integrated differential width is evaluated by approximating the integration kernel via Chebyshev polynomials, where individual terms of the polynomial expansion are made related to the Euclidean correlator, leading to a deterministic way to estimate the observable. This method is interpreted as a certain extension of the reconstruction for the spectrum function in previous works (see, {\it e.g.,} Ref.~\cite{Asakawa:2000tr}), where the smearing procedure is additionally involved. For the numerical inputs in (c), however, duality-violating components that arise from the finite-distance singularity fall off quickly in the limit where the bottom quark is sufficiently heavy so that whether the obtained prediction in the Minkowski domain truly reflects duality-violating components seems rather nontrivial. Meanwhile, the fully integrated width predicted in this method gives a result consistent with the OPE \cite{Gambino:2020crt}.
\par
In this work, in the presence of the right-handed current (RHC), we show that a duality-violating contribution has some aspects different from the one in the standard model (SM), on the basis of the instanton-based approach established in Ref.~\cite{Chibisov:1996wf}. In particular, the fermionic zero mode \cite{tHooft:1976snw} for the intermediate quark in $\bar{B}\to X_ul\nu$ decays gives a nonvanishing contribution for the RHC, whereas this term is absent in the SM for the single instanton since it is prohibited by its chirality structure. Within a certain approximation, we find that a duality-violating contribution that arises from the finite-distance singularity for the RHC is enhanced relative to the one in the SM typically by $\sim m_b/m^*$, where $m^*$ is an effective mass of the light quark discussed later. Furthermore, it is shown that the lepton energy distribution exhibits a parametrically larger behavior than the total width, analogous to the nonzero-mode contribution in the SM \cite{Chibisov:1996wf}, indicating that the duality-violating contribution from the finite-distance singularity to the differential width is comparable to the chirally suppressed perturbative contribution in the RHC for the $\bar{B}$ decays, while the same contribution to the $D$ decays is even larger up to uncertainty in the saddle point approximation.
\par
This paper is organized as follows: In Sec.~\ref{Sec:II}, the formalism for the inclusive semileptonic decays of the $\bar{B}$ meson is given in the presence of the RHC. Subsequently, a duality-violating contribution induced by an instanton is discussed, with a particular emphasis on the zero mode for the up-quark propagator in Sec.~\ref{Sec:III}. Furthermore, numerical results associated with the (differential) width are presented for $\bar{B}\to X_ul\nu$ and also for $D\to X_dl\nu$. Finally, a concluding remark is given in Sec.~\ref{Sec:IV}.
\section{Width in $\bar{B}\to X_ul\nu$ decay}
\label{Sec:II}
In the presence of the RHC for quarks, the effective Hamiltonian that leads to $b\to ul\nu$ decay is written as
\bea
\mathcal{H}_{\rm eff}=\frac{4G_F}{\sqrt{2}}V_{ub}(c_{\rm L}J_{\rm L}^\mu \mathcal{J}_{\mathrm{L}\mu}+c_{\rm R}J_{\rm R}^\mu \mathcal{J}_{\mathrm{L}\mu})+\mathrm{H.c.}\label{Eq:1}
\eea
In Eq.~(\ref{Eq:1}), $J_{\rm X}^\mu=\bar{u}\gamma^\mu P_{\rm X}b~(\mathrm{X}=\mathrm{L}, \mathrm{R})$ and $\mathcal{J}_{\rm L}^\mu=\bar{l}\gamma^\mu P_{\rm L}\nu$ are quark and lepton currents, respectively. In the SM, $c_{\rm L}=1$ and $c_{\rm R}=0$ are realized. We introduce the following forward scattering matrix element:
\bea
T^{\mu\nu}&=&c_{\rm L}^2T^{\mu\nu}_{\rm LL}+c_{\rm L}c_{\rm R}T^{\mu\nu}_{\rm LR}+c_{\rm R}c_{\rm L}T^{\mu\nu}_{\rm RL}+c_{\rm R}^2T^{\mu\nu}_{\rm RR},\label{Eq:Wdec}\\
T^{\mu\nu}_{\rm XY}&=&-i\int\mathrm{d}^4xe^{-iq\cdot x}\bra{\bar{B}_v}T[J^{\mu\dagger}_{\rm X}(x)J^{\nu}_{\rm Y}(0)]\ket{\bar{B}_v},
\eea
where $T^{\mu\nu}$ represents a total contribution while $T^{\mu\nu}_{\rm XY}$ gives an individual term specified by chirality indices. With Eq.~(\ref{Eq:Wdec}), the triple differential width is represented as
\bea
\frac{\mathrm{d}^3\Gamma}{\mathrm{d} E_l \mathrm{d} E_\nu \mathrm{d}q^2}&=&-\frac{G_F^2|V_{ub}|^2}{4\pi^4}\mathrm{Im}(T_{\mu\nu}L^{\mu\nu}),\label{Eq:tri}\\
L^{\mu\nu}&=&2(p_l^\mu p_{\bar{\nu}}^\nu + p_{\bar{\nu}}^\mu p_l^\nu -g^{\mu\nu}p_l\cdot p_{\bar{\nu}}-i\epsilon^{\mu\nu\rho\sigma} p_{l\rho} p_{\bar{\nu}\sigma}).
\eea
In Eq.~(\ref{Eq:tri}), $E_l~(E_\nu)$ is the charged lepton (neutrino) energy while $q$ gives a four-momentum of the lepton system, all of which are defined at the rest frame of the $\bar{B}$ meson. Similarly to Eq.~(\ref{Eq:Wdec}), the lepton energy distribution of the semileptonic decay is given by
\bea
\frac{\mathrm{d}\Gamma}{\mathrm{d}y}=
c_{\rm L}^2\left.\frac{\mathrm{d}\Gamma}{\mathrm{d}y}\right|_{\rm LL}+
c_{\rm L}c_{\rm R}\left.\frac{\mathrm{d}\Gamma}{\mathrm{d}y}\right|_{\rm LR}+
c_{\rm R}c_{\rm L}\left.\frac{\mathrm{d}\Gamma}{\mathrm{d}y}\right|_{\rm RL}+
c_{\rm R}^2\left.\frac{\mathrm{d}\Gamma}{\mathrm{d}y}\right|_{\rm RR},\label{Eq:ydist}
\eea
where $y=2E_l/m_b$ is the dimensionless kinematical variable. In what follows, the second and third terms in Eq.~(\ref{Eq:ydist}) are referred to as the RHC contributions. As for the last term in Eq.~(\ref{Eq:ydist}), it has a structure identical to one in the SM up to chirality flipping. Henceforth, the discussion for the last term in Eq.~(\ref{Eq:ydist}) is omitted in the remaining part of this paper. 
\par
The contribution of the SM in Eq.~(\ref{Eq:ydist}) is well known and obtained in, {\it e.g.}, Refs.~\cite{Chay:1990da, Bigi:1992su, Bigi:1993fe,Blok:1993va,Manohar:1993qn,Mannel:1993su,Neubert:1993ch,Falk:1993vb}. As to the RHC contributions, the differential rate corresponding to the $\bar{B}\to X_cl\nu$ decay is calculated in Ref.~\cite{Dassinger:2007pj}. With $r=(m_u/m_b)^2$, the lepton energy distributions calculated in the leading perturbative analysis are
\bea
\left.\frac{1}{\Gamma_0}\frac{\mathrm{d}\Gamma}{\mathrm{d} y}\right|_{\rm LL}&=&2y^2(3-2y)-6y^2r - \frac{6y^2r^2}{(1-y)^2}+\frac{2y^2(3-y)r^3}{(1-y)^3},\label{Eq:lepLL}\\
\left.\frac{1}{\Gamma_0}\frac{\mathrm{d}\Gamma}{\mathrm{d} y}\right|_{\rm LR}&=&\left.\frac{1}{\Gamma_0}\frac{\mathrm{d}\Gamma}{\mathrm{d} y}\right|_{\rm RL}=-6\sqrt{r}y^2\left(1-\frac{r}{1-y}\right)^2,\label{Eq:lepLR}
\eea
both of which are normalized by $\Gamma_0=G_F^2m_b^5|V_{ub}|^2/(192\pi^3)$.  In obtaining the above expressions, we terminated the phase space integral via the partonic delta function and also used formulas for the matrix elements given by $\bra{\bar{B}_v}\bar{b}_v\gamma^\mu b_v\ket{\bar{B}_v}=v^\mu$ and $\bra{\bar{B}_v}\bar{b}_v\gamma^\mu\gamma^\nu b_v\ket{\bar{B}_v}=g^{\mu\nu}$, which follow from the heavy quark symmetry \cite{Isgur:1989vq}. By further integrating Eqs.~(\ref{Eq:lepLL}) and (\ref{Eq:lepLR}), one can also obtain the total widths for the semileptonic decay:
\bea
\left.\frac{\Gamma}{\Gamma_0}\right|_{\mathrm{LL}}^{\rm pert}&=&1-8r-12r^2\ln r+8r^3-r^4,\label{Eq:LLpartonic}\\
\left.\frac{\Gamma}{\Gamma_0}\right|^{\rm pert}_{\rm LR}&=&\left.\frac{\Gamma}{\Gamma_0}\right|^{\rm pert}_{\rm RL}=-2\sqrt{r}[1+3r(3+2\ln r)-3r^2(3-2\ln r)-r^3].\label{Eq:LRpartonic}
\eea
The results corresponding to $D\to X_dl\nu$ can be obtained by replacing $m_b\to m_c,~ m_u\to m_d$ in Eqs.~(\ref{Eq:lepLL})-(\ref{Eq:LRpartonic}). It is evident that the leading perturbative contribution for the RHC in Eqs.~(\ref{Eq:lepLR}) and (\ref{Eq:LRpartonic}) is suppressed by the light quark mass, since it is chirally disfavored. 
\par
An instanton-induced correction to the SM part is evaluated by means of the dilute gas approximation for $B\to X_ue\nu$ \cite{Chay:1994si, Falk:1995yc} and for $B\to X_s\gamma$ \cite{Chay:1994dk} decays. Meanwhile, duality violation that arises from finite-distance singularity of the light quark Green function for the SM part is investigated in Ref.~\cite{Chibisov:1996wf} via the saddle point approximation. These previous works are all based on the nonzero-mode contributions in the quark propagator. In the next section, a duality-violating correction to the RHC contribution in Eqs.~(\ref{Eq:lepLR}) and (\ref{Eq:LRpartonic}) from the zero mode is calculated.
\section{Zero-mode and RHC contribution}
\label{Sec:III}
In order to analyze the instanton-induced contribution to the fully integrated width for $\bar{B}\to X_ul\nu$ decay, we consider formulas alternative to the one based on momentum space integration discussed in the previous section:
\bea
\Gamma_{\rm XY}&=&-16G_F^2|V_{ub}|^2\mathrm{Im}(\tilde{\mathcal{T}}_{\rm XY}),\label{Eq:totalDef}\\
\tilde{\mathcal{T}}_{\rm XY}&=&
-i\int\mathrm{d}^4x \bra{\bar{B}_v}J_{\rm X}^{\dagger \mu}(x)J_{\rm Y}^{\nu}(0) \ket{\bar{B}_v}\tilde{L}_{\mu\nu}(x),\label{Eq:Txyten}\\
\tilde{L}_{\mu\nu}(x)&=&-\frac{1}{2\pi^4}\frac{1}{x^8}(2x_\mu x_\nu-x^2g_{\mu\nu}),\label{Eq:leptspa}
\eea
where the leptonic tensor in Eq.~(\ref{Eq:leptspa}) is defined in the position space. In what follows, we use the Euclidean notation unless otherwise specified. In the singular gauge, the instanton configuration is given by $(D_\mu=\partial_\mu-iA_\mu^a \tau^a/2)$
\bea
A_{\mu}^a=\bar{\eta}_{a\mu\nu}\frac{(x-z)_\nu\rho^2}{(x-z)^2[(x-z)^2+\rho^2]},\label{Eq:instsing}
\eea
where $\rho$ and $z$ are the size and center of the instanton, respectively, while $\bar{\eta}_{a\mu\nu}$ is a conventional 't Hooft symbol. The fermion Green function in the presence of the instanton is calculated in Refs.~\cite{Brown:1977eb, Brown:1978bta, Carlitz:1978xu, Andrei:1978xg,Lee:1979sm}. In particular, the propagator of the up quark is
\bea
S(x, y, z)=\displaystyle\sum_{n=0}^\infty\frac{\psi_n(x)\psi_n^\dagger(y)}{\lambda_n-m_u}=S_{\rm zm}+S_{\rm nzm}+\mathcal{O}(m_u),
\eea
where $S_{\rm zm}$ and $S_{\rm nzm}$ are the contributions that arise from the zero mode $(n=0)$ \cite{Levine:1978ge, Diakonov:1979nj} and the nonzero mode $(n\neq 0)$ \cite{Andrei:1978xg, Levine:1978ge,
Brown:1978bta, Brown:1977eb, Schafer:2004ke}, respectively:
\bea
S_{\rm zm}&=&-\frac{\rho^2}{8\pi^2m_u}\frac{\slashed{\tilde{x}}\gamma_\mu\gamma_\nu\slashed{\tilde{y}}}{\sqrt{\tilde{x}^2\tilde{y}^2}(\tilde{x}^2+\rho^2)^{3/2}(\tilde{y}^2+\rho^2)^{3/2}}\tau^-_\mu\tau^+_\nu P_{\rm L},\label{Eq:zeroProp}\\
S_{\rm nzm}&=&-\frac{\slashed{\Delta}}{2\pi^2 \Delta^4}\frac{\sqrt{\tilde{x}^2\tilde{y}^2}}{\sqrt{\tilde{x}^2+\rho^2}\sqrt{\tilde{y}^2+\rho^2}}\left[1+\frac{\rho^2}{\tilde{x}^2\tilde{y}^2}(\tau^-\cdot \tilde{x})(\tau^+\cdot \tilde{y})\right]-\frac{i}{4\pi^2\tilde{x}^2\tilde{y}^2\Delta^2}\frac{\sqrt{\tilde{x}^2\tilde{y}^2}}{\sqrt{\tilde{x}^2+\rho^2}\sqrt{\tilde{y}^2+\rho^2}}\nn\\
&&\times
(\tau^-\cdot \tilde{x})\left[\frac{\rho^2}{\tilde{x}^2+\rho^2}\slashed{\tau}^+(\tau^-\cdot\Delta) P_{\rm R}+\frac{\rho^2}{\tilde{y}^2+\rho^2}(\tau^+\cdot\Delta)\slashed{\tau}^- P_{\rm L}
\right](\tau^+\cdot \tilde{y}),\label{Eq:zeroProp2}
\eea
for $\Delta = x-y$, $\tilde{x}=x-z$, and $\tilde{y}=y-z$. In the above relations, $\tau^\pm_\mu$ is a matrix that resides in the $2\times 2$ subspace of the SU(3) group:
\bea
\tau^\pm_\mu&=&(\vec{\tau}, \mp i), \quad \tau_{\mu}^-\tau^+_{\nu} =\delta_{\mu\nu}+\bar{\eta}_{\mu\nu a}\tau_a,
\eea
with $\vec{\tau}$ being the Pauli matrix. Equations (\ref{Eq:zeroProp}) and (\ref{Eq:zeroProp2}) are understood as cases of the trivial color orientation ($U=1$) for the instanton, which are to be covariantly transformed together with the configuration in Eq.~(\ref{Eq:instsing}). The propagators in the presence of an anti-instanton can be obtained via the interchanges of $\tau^-\leftrightarrow \tau^+$ and $\mathrm{L}\leftrightarrow\mathrm{R}$ in Eqs.~(\ref{Eq:zeroProp}) and (\ref{Eq:zeroProp2}).\par
In order to calculate the zero-mode contribution to the width, we make use of the single instanton approximation (SIA), which has been applied to the analysis of the hadronic correlator, {\it etc}. At a sufficiently short-distance region, the correlator in this method is evaluated with the single instanton given that the separation between instantons is much longer. Effects of the multi-instanton, responsible for the chiral symmetry breaking \cite{Diakonov:1979nj}, are taken into account via replacement of the current quark mass $(m_u)$ with the effective mass $(m^*)$ in Eq.~(\ref{Eq:zeroProp}). The advantage of the SIA is that it offers an analytical investigation of the currently considered duality violation based on the finite-distance singularity.  In order for this approximation to work properly, the same choice of $m^*$ should universally work for various observables while, at the same time, the distance for the correlator should be small enough. The validity of the SIA is investigated in Ref.~\cite{Faccioli:2001ug}: It revealed that for correlators which include double contributions of the zero mode, such as the four-quark condensate and the pion correlator, the SIA successfully works. Indeed, for $|x|\lessapprox 0.6~\mathrm{fm}$, the pion correlator is well reproduced by the leading single instanton \cite{Faccioli:2001ug}. However, for the two-quark condensate, which includes the contribution of a single zero mode, the full instanton liquid gives an effective mass different from one for the four-quark condensate so that an uncertainty is involved in regards to universality. In viewing this issue, we consider $120~\mathrm{MeV}\leq m^*\leq 177~\mathrm{MeV}$ as a typical uncertainty range, where the lower and upper limits come from the values based on the two-quark condensate in the random instanton liquid model (RILM) and the interacting instanton liquid model (IILM), respectively \cite{Faccioli:2001ug}.
\par
In the SIA, one can find that the zero mode does not contribute to $\tilde{\mathcal{T}}_{\rm LL}$ while the nonzero mode does not contribute to $\tilde{\mathcal{T}}_{\rm XY}$ for $(\mathrm{X}, \mathrm{Y})=(\mathrm{L}, \mathrm{R}), (\mathrm{R}, \mathrm{L})$, since those are chirally disfavored. Thus, $\tilde{\mathcal{T}}_{\rm LL}$ receives only the nonzero-mode contribution, while the zero mode from the (anti-)instanton gives a nonvanishing correction to $\tilde{\mathcal{T}}_{\rm LR}$ $(\tilde{\mathcal{T}}_{\rm RL})$. In what follows, we analyze the instanton-induced contribution to $\tilde{\mathcal{T}}_{\rm LR}$, since the anti-instanton correction to $\tilde{\mathcal{T}}_{\rm RL}$ is essentially similar to the former up to some replacement of indices.
\par
The zero-mode contribution that originates from the RHC to Eq.~(\ref{Eq:Txyten}) is
\bea
\tilde{\mathcal{T}}_{\rm LR}^{\rm zm}&=&i\int\frac{\mathrm{d}\rho}{\rho^5}d(\rho)\mathrm{d}U\mathrm{d}^4z\mathrm{d}^4xe^{iQ\cdot x} \bra{\bar{B}_v}\bar{b}_v(x)\gamma_\mu P_{\rm L}S_{\rm zm}(x, 0, z)\gamma_\nu P_{\rm R}
b_v(0)\ket{\bar{B}_v}\tilde{L}_{\mu\nu}(x),\label{Eq:instDef}
\eea
with $m_bv_0=iQ_4$ and $x_0=-ix_4$, where $Q_4$ is assumed to scale like $m_b$. 
In Eq.~(\ref{Eq:instDef}), $d(\rho)$ is the dimensionless size distribution of an instanton simply taken as $d(\rho)=d_0\rho \delta (\rho-\rho_0)$. In Eq.~(\ref{Eq:instDef}), the notation of the bottom quark field is kept in the Minkowski domain. It should be noted that the propagator for the zero mode in Eq.~(\ref{Eq:zeroProp}) is rewritten by
\bea
S_{\rm zm}(x, y, z)&=&-\frac{1}{m^*}\frac{\rho^2}{2\pi^2}F(x, y, z)P_{\rm L}+(\mathrm{color~nonsinglet}),\label{Eq:zerosim}\\
F(x, y, z)&=&\frac{\tilde{x}\cdot\tilde{y}}{\sqrt{\tilde{x}^2\tilde{y}^2}(\tilde{x}^2+\rho^2)^{3/2}(\tilde{y}^2+\rho^2)^{3/2}},
\eea
where the $m_u$ in Eq.~(\ref{Eq:zeroProp}) is replaced by $m^*$ for the SIA. The color nonsinglet part in Eq.~(\ref{Eq:zerosim}) vanishes when one takes an average over the color orientation of the instanton and will be omitted in the following discussion. By substituting Eq.~(\ref{Eq:zerosim}) into Eq.~(\ref{Eq:instDef}) and performing the integral with respect to the instanton size (this does not affect the result as long as we consider the fixed-sized instanton), one can obtain
\bea
\tilde{\mathcal{T}}_{\rm LR}^{\rm zm}&=&-\frac{id_0}{2\pi^2\rho^2m^*} \int\mathrm{d}^4x\mathrm{d}^4z\mathrm{d}UF(x, 0, z)e^{iQ\cdot x}\bra{\bar{B}_v}\bar{b}_v(x)\gamma_\mu \gamma_\nu P_{\rm R}b_v(0)\ket{\bar{B}_v}\tilde{L}_{\mu\nu}(x).\label{Eq:mid}
\eea
By implementing the Feynman parametrization, we combine the denominators in $F(x,  0, z)$:
\bea
F(x, 0, z)e^{iQ\cdot x}&=&\int_0^1\mathrm{d}\alpha 
f(x_4, \alpha, \vec{x}, z),\\
f(x_4, \alpha, \vec{x}, z)&=&-\frac{8}{\pi}\frac{[\alpha(1-\alpha)]^{-\frac{5}{2}}}{(x_4^2+\bar{x}^2)^3}\frac{\tilde{x}\cdot z}{\sqrt{\tilde{x}^2} \sqrt{z^2}}
e^{iQ\cdot x},\\
\bar{x}&=&\sqrt{\frac{(z-\alpha x)^2+\rho^2_0}{\alpha(1-\alpha)}+\vec{x}^2}.
\eea
In order to capture duality violation that arises from the finite-distance singularity, we terminate the integration of $x_4$ by closing the contour via the upper semicircle on the complex plane and then picking up the residue at $x_4=i\bar{x}$:
\bea
\label{Eq:integ4}
\int_{-\infty}^{+\infty}\mathrm{d}x_4f(x_4, \alpha, \vec{x}, z)
&\simeq&-\frac{4Q_4^2}{\rho^3_0}\bar{f}(\alpha, \vec{x}, z)e^{-Q_4 \bar{x}},\\
\bar{f}(\alpha, \vec{x}, z)&=&\frac{[\alpha(1-\alpha)]^{-\frac{5}{2}}}{32}\left(\frac{2\rho_0}{\bar{x}}\right)^3\frac{\tilde{x}\cdot z}{\sqrt{\tilde{x}^2} \sqrt{z^2}}e^{-i\vec{Q}\cdot \vec{x}},
\eea
where the term that has the largest power of $Q_4$ is extracted in Eq.~(\ref{Eq:integ4}).
\par
Here, we consider the limit where $m_b$ is sufficiently larger than an inverse of the instanton size, which is a reasonable assumption for the typical value supported by the instanton liquid model \cite{Shuryak:1981ff}. In this case, the integrals of $\alpha$, $\vec{x}$, and $z$ are nearly Gaussian that are sharply peaked with narrow intervals:
\bea
\vec{x}^2\sim \frac{\rho_0}{m_b},\quad
\left(\alpha-\frac{1}{2}\right)^2\sim \frac{1}{m_b\rho_0},\quad \left(z-\frac{x}{2}\right)^2\sim \frac{\rho_0}{m_b}.
\eea
Aside from the exponential of $e^{-Q_4 \bar{x}}$ in Eq.~(\ref{Eq:integ4}), all the remaining prefactors including the leptonic tensor in the position space are evaluated at the saddle point. For $\bar{b}_v(x)$ in Eq.~(\ref{Eq:mid}), one needs to solve the equation of motion to obtain the heavy quark propagation in the medium of the instanton, leading to $b_v(x)={\cal U}(x)b(0, \vec{x})+\mathcal{O}(1/m_b\rho_0)$ \cite{Chibisov:1996wf}. In the limit of large $m_b\rho_0$, the propagation matrix is approximately equal to unity, and, therefore, the bottom quark bilinear is given by the local operator at the saddle point, in which case the average over color orientation is taken straightforwardly, leading to
\bea
\tilde{\mathcal{T}}_{\rm LR}^{\rm zm}&=&i\frac{2}{3}\frac{d_0Q_4^2}{\pi^2\rho^5_0m^*}\int\mathrm{d}\alpha\mathrm{d}^3x\mathrm{d}^4ze^{-Q_4 \bar{x}}\bra{\bar{B}_v}\bar{b}_v\gamma_\mu \gamma_\nu b_v\ket{\bar{B}_v}\tilde{L}_{\mu\nu}(x).\label{Eq:eucres}
\eea
One can subsequently perform the Gaussian integral via $\int \mathrm{d}\alpha\mathrm{d}^3x\mathrm{d}^4ze^{-Q_4\bar{x}}\simeq 4\pi^4\rho^3_0e^{-2Q_4\rho_0}/Q_4^4$ and obtain the result of the total width in Eq.~(\ref{Eq:totalDef}) by Wick rotating the quantity in Eq.~(\ref{Eq:eucres}) back to the Minkowski domain:
\bea
\left.\frac{\Gamma}{\Gamma_0}\right|^{\rm zm}_{\rm LR}=-\frac{2}{3}d_0\left(\frac{m_b}{m^*}\right)\frac{192\pi }{(m_b\rho_0)^8}\cos(2m_b\rho_0).\label{Eq:RHCres}
\eea
The RHC contribution in Eq.~(\ref{Eq:RHCres}) is to be contrasted with one in the SM based on the nonzero mode \cite{Chibisov:1996wf}:
\bea
\left.\frac{\Gamma}{\Gamma_0}\right|^{\rm nzm}_{\rm LL}=-\frac{2}{3}d_0\frac{96\pi }{(m_b\rho_0)^8}\sin(2m_b\rho_0).\label{Eq:ShifLL}
\eea
The results for $D\to X_dl\nu$ can be obtained by replacing $m_b\to m_c$ in Eqs.~(\ref{Eq:RHCres}) and (\ref{Eq:ShifLL}).
\par
Before addressing the numerical result, we note the following aspects: In Ref.~\cite{Chibisov:1996wf}, the instanton-based approach is regarded as a {\it model} of duality violation, yet the fixed-sized single instanton is certainly insufficient to be a sole dominant ingredient in the QCD vacuum and, therefore, leading to some obvious drawbacks. In order for the model to contribute to the phenomenology in $e^+e^-\to\mathrm{hadrons}$, hadronic tau decay, and $D$ meson semileptonic decay, a value of the instanton density (size) is larger (smaller) than what is indicated by the instanton liquid model. In addition, for the same choice of the parameters, the predicted values for the oscillation period for the $R$ ratio as well as the gluon condensate are too large.
\par
In this work, viewing the aforementioned incapability of the single instanton, we do not consider this approach as a model in which the instanton size and density are regarded as free parameters. Instead, those quantities are fixed to the values suggested by the instanton liquid model. Specifically, we set $\rho_0=1/3~\mathrm{fm}$ and $n=1~\mathrm{fm}^{-4}$, corresponding to $d_0=n\rho_0^4/2=6.2\times 10^{-3}$, much smaller than $d_0=9\times 10^{-2}$ adopted in Ref.~\cite{Chibisov:1996wf}.  As for the quark masses, the pole mass and $\overline{\rm MS}$ masses are considered for heavy quarks, while the light quark masses in the perturbative expressions in Eqs.~(\ref{Eq:LLpartonic}) and (\ref{Eq:LRpartonic}) are taken as $\overline{\rm MS}$ masses at the scale of the associated heavy quark mass. As described before, the effective mass of the light quark is varied for the values obtained in the simulations of the RILM and the IILM, on the basis of the quark condensate.
\par
In order to investigate a typical size of duality violation for the total width, we consider the following object:
\bea
R_{\rm XY}^{\rm tot}[H_Q\to X_ql\nu]=\left|\frac{\tilde{\Gamma}^{\rm I}_{\rm XY}}{\Gamma^{\rm pert}_{\rm XY}}\right|,\label{Eq:totr}
\eea
for $(H_Q, q)=(\bar{B}, u)$ and $(D, d)$. In Eq.~(\ref{Eq:totr}), ${\rm(X, Y, I)=(L, L, nzm)}$ corresponds to the result for SM, while ${\rm(X, Y, I)=(L, R, zm)}$ is associated with the RHC contribution. The denominator ($\Gamma^{\rm pert}_{\rm XY}$) in Eq.~(\ref{Eq:totr}) is extracted from Eqs.~(\ref{Eq:LLpartonic}) and (\ref{Eq:LRpartonic}), while for the numerator ($\tilde{\Gamma}^{\rm I}_{\rm XY}$), the expressions in Eqs.~(\ref{Eq:RHCres}) and (\ref{Eq:ShifLL}) in which the oscillating coefficients are removed are used. This is because  duality violation may vanish if the heavy quark mass is accidentally located at oscillating nodes.
\par
The numerical results for the total semileptonic widths for $\bar{B}$ and $D$ decays are shown in Tables~\ref{Tab:1} and \ref{Tab:2}, respectively. By comparing the two tables, one can find that the considered duality-violating contributions are generically larger for $D$ decays than $\bar{B}$ decays. By further taking the ratio of the form $R_{\rm LR}^{\rm tot}[H_Q\to X_ql\nu]/R_{\rm LL}^{\rm tot}[H_Q\to X_ql\nu]$ for $(H_Q, q)=(\bar{B}, u), (D, d)$, we find that the importance of the duality-violating contribution for the RHC is larger than that of the SM by from $5\times 10^4$ to $1\times 10^5$ for $\bar{B}$ decays and from $2\times 10^3$ to $5\times 10^3$ for $D$ decays. Notwithstanding this huge relative importance, the absolute size of duality violation ($R_{\rm LR}^{\rm tot}$) is small, especially for $\bar{B}$ decays in Table~\ref{Tab:1}. As we shall see later, for the differential decay rate, the zero-mode-induced duality violation can be as large as the chirally suppressed perturbative contribution for $\bar{B}$ and $D$ decays, depending on the phase space region, since the smearing procedure potentially gives an excessive suppression of duality violation.
\begin{table}
\begin{center}
\caption{Ratio of instanton-induced duality violation to corresponding perturbative contribution in $\bar{B}\to X_ul\nu$ decays. The first row shows the adopted values of the bottom quark mass and effective mass for the light quark in GeV unit. The second and third rows show the ratios for the RHC and the SM, respectively.}
\label{Tab:1}
\begin{tabular}{ ccccc} 
\hline\hline
  $(m_b, m^*)\: [\mathrm{GeV}]$& $(4.18, 0.120)$ & $(4.78, 0.120)$ & $(4.18, 0.177)$ & $(4.78, 0.177)$\\ \hline
$R_{\rm LR}^{\rm tot}[\bar{B}\to X_ul\nu]$   & $2\times 10^{-2}$ & $7\times 10^{-3}$ & $1\times 10^{-2}$   & $5\times 10^{-3}$  \\ 
$R_{\rm LL}^{\rm tot}[\bar{B}\to X_ul\nu]$ & $2\times 10^{-7}$ & $7\times 10^{-8}$ & $2\times 10^{-7}$ &  $7\times 10^{-8}$ \\ 
\hline\hline
\end{tabular}
\end{center}
\end{table}
\begin{table}
\begin{center}
\caption{The same table as Table~\ref{Tab:1} except that $\bar{B}$ decays are properly replaced by $D$ decays.}
\label{Tab:2}
\begin{tabular}{ccccc} 
\hline\hline
  $(m_c, m^*)\: [\mathrm{GeV}]$& $(1.27, 0.120)$ & $(1.67, 0.120)$ & $(1.27, 0.177)$ & $(1.67, 0.177)$\\ \hline
$R_{\rm  LR}^{\rm tot}[D\to X_dl\nu]$   & $7$ & $1$ & $5$   & $1$  \\ 
$R_{\rm  LL}^{\rm tot}[D\to X_dl\nu]$ & $3\times 10^{-3}$ & $3\times 10^{-4}$ & $3\times 10^{-3}$ &  $3\times 10^{-4}$ \\ 
\hline\hline
\end{tabular}
\end{center}
\end{table}
\par
In the previous work \cite{Chibisov:1996wf}, it has been found that duality violation in the SM that arises from the nonzero mode gives a parametrically larger contribution to the differential width than the total width receives. This paradoxical behavior has been interpreted as follows: Around the end point region of the phase space, the theoretical expression for the differential width, which relies on the semiclassical approximation, is not valid {\it per se}. Meanwhile, in order for the duality-violating term to satisfy the dispersion relation, the OPE-like counterterm, which originates from the singularity at $x^2=0$, must be accompanied. With the dispersion relation properly realized due to this treatment, the phase space domain of the lepton momenta can be deformed as an unphysical region. In this case, one can obtain the total width by integrating the differential width that ensures an entire consistency.
\par
For the duality-violating contribution that arises from the zero mode for the RHC, one also finds a paradoxical behavior similar to the SM for the differential width. One difference between the SM and the RHC in this regard is that, for the RHC, the zero-mode contribution leads to the singularity at $x^2=0$ (obviously in the case of $y=0$), whereas the zero-mode contribution is prohibited due to the chirality structure in the SIA. Hence, the consistent realization of the dispersion relation for the RHC would require the nontrivial contribution from the zero mode in the short-distance expansion, in order to guarantee that the integration of the differential width consistently gives the total width.
\par
As a next step, we investigate the lepton energy distributions for $H_Q\to X_ql\nu$ decays. At this stage, care must be taken for the aforementioned apparent paradoxical aspect for the width. Otherwise, a naive integration over the neutrino momentum to obtain the lepton energy distribution could easily overestimate the observable. Since the consistent treatment over the end point region requires the {\it ad hoc} OPE-like counterterms as discussed before, in this work we do not explicitly integrate over the neutrino momentum. To this end, the propagators for neutrino and up quark are considered in the position space, while the charged lepton propagator is expanded in the momentum space. In practice, to analyze the differential width in the presence of an instanton, a procedure similar to the total width can be implemented; we replace $q=p_l+p_\nu$ by $q=p_l$ and also the leptonic tensor by an appropriate one. That is, we use the following formulas for the differential width (for definitiveness, the formulas for $\bar{B}$ decays are considered below):
\bea
\mathrm{d}\Gamma_{\rm XY}&=&-16G_F^2|V_{ub}|^2\mathrm{Im}\left(\bar{\cal T}_{\rm XY}\right)\frac{\mathrm{d}^3p_l}{(2\pi)^32E_l},\label{Eq:integratedlep}\\
\bar{\cal T}_{\rm XY}&=&-i\int\mathrm{d}^4xe^{-ip_l\cdot x}\bra{\bar{B}_v}T[J_{\rm X}^{\dagger \mu} (x)J_{\rm Y}^\nu(0)]\ket{\bar{B}_v}\bar{L}_{\mu\nu},\label{Eq:XYlep}\\
\bar{L}_{\mu\nu}&=&-\frac{i}{\pi^2 x^4}(x_\mu p_{l\nu} +p_{l\mu} x_\nu-x\cdot p_lg_{\mu\nu}+i\epsilon_{\mu\nu\rho\sigma}x^\rho p^\sigma_l).\label{Eq:lep_ener_ten}
\eea
It is straightforward to confirm that the partonic result for the SM in Eq.~(\ref{Eq:lepLL}) can be reproduced via Eq.~(\ref{Eq:integratedlep}) for the case where the free propagator is used for the up quark in Eq.~(\ref{Eq:XYlep}). Duality violation for both RHC and SM can be calculated on the basis of Eq.~(\ref{Eq:integratedlep}). As a result, the normalized lepton energy distributions for the RHC and the SM are,
\bea
\left.\frac{1}{\Gamma_0}\frac{\mathrm{d}\Gamma}{\mathrm{d}y}\right|_{\rm LR}^{\rm zm}&=&-\frac{2}{3}d_0\left(\frac{m_b}{m^*}\right)\frac{96\pi}{(m_b\rho_0)^5}\frac{y^2}{\left(1-\displaystyle\frac{y}{2}\right)^2}\sin\left(2m_b\rho_0\left(1-\displaystyle\frac{y}{2}\right)\right),\label{Eq:upperone}\\
\left.\frac{1}{\Gamma_0}\frac{\mathrm{d}\Gamma}{\mathrm{d}y}\right|_{\rm LL}^{\rm nzm}&=&-\frac{2}{3}d_0\frac{96\pi}{(m_b\rho_0)^5}\frac{y^2}{\left(1-\displaystyle\frac{y}{2}\right)^3}\cos\left(2m_b\rho_0\left(1-\displaystyle\frac{y}{2}\right)\right).\label{Eq:lepSM}
\eea
It should be noted that the above results are obtained without imposing the dispersion relation, which is different from the method in Ref.~\cite{Chibisov:1996wf}. Nonetheless, the proportionality to $(m_b\rho_0)^{-5}$ in Eq.~(\ref{Eq:lepSM}) is consistent with Ref.~\cite{Chibisov:1996wf}, since the neutrino propagator $(\propto x_\mu/x^4)$ is simply fixed by $x^2=-4\rho^2_0$ in Eq.~(\ref{Eq:lep_ener_ten}) while the nontrivial realization of the dispersion relation in Ref.~\cite{Chibisov:1996wf} leads to the same dependence.
\par
\begin{figure}[H]
\begin{center}
\includegraphics[scale=0.85]{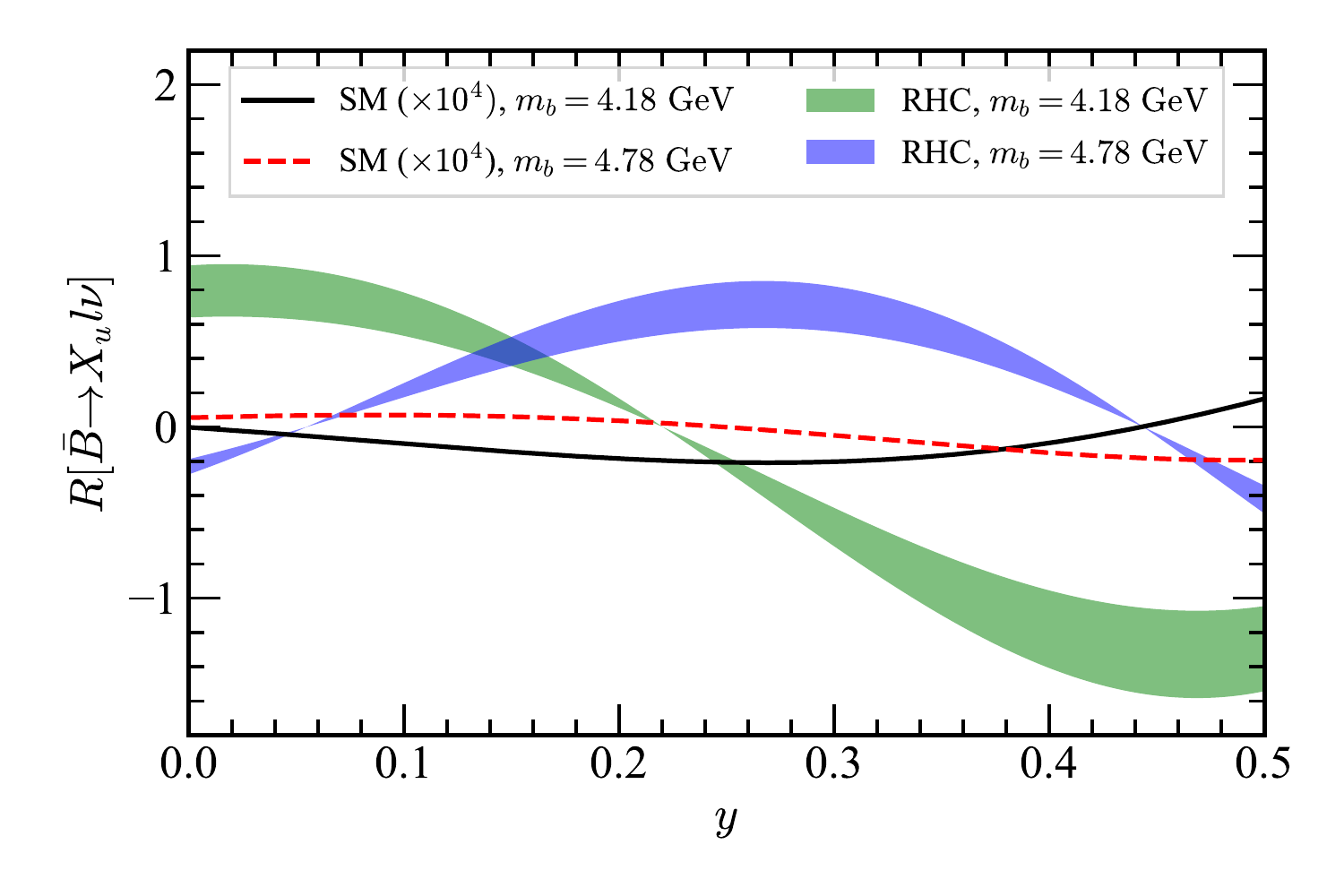}
\vspace{-10mm}
\caption{Ratio of duality-violating component to corresponding perturbative contribution for lepton energy distribution in $\bar{B}\to X_ul\nu$ decays. The black solid (red dashed) line represents the result of the SM multiplied by $10^4$ for $m_b=4.18~(4.78)~\mathrm{GeV}$. The green (blue) band stands for the result for the RHC obtained by varying $120~\mathrm{MeV}\leq m^*\leq 177~\mathrm{MeV}$ with $m_b=4.18~(4.78)~\mathrm{GeV}$. }
\label{Fig:1}
\end{center}
\end{figure}
\begin{figure}[H]
\begin{center}
\includegraphics[scale=0.85]{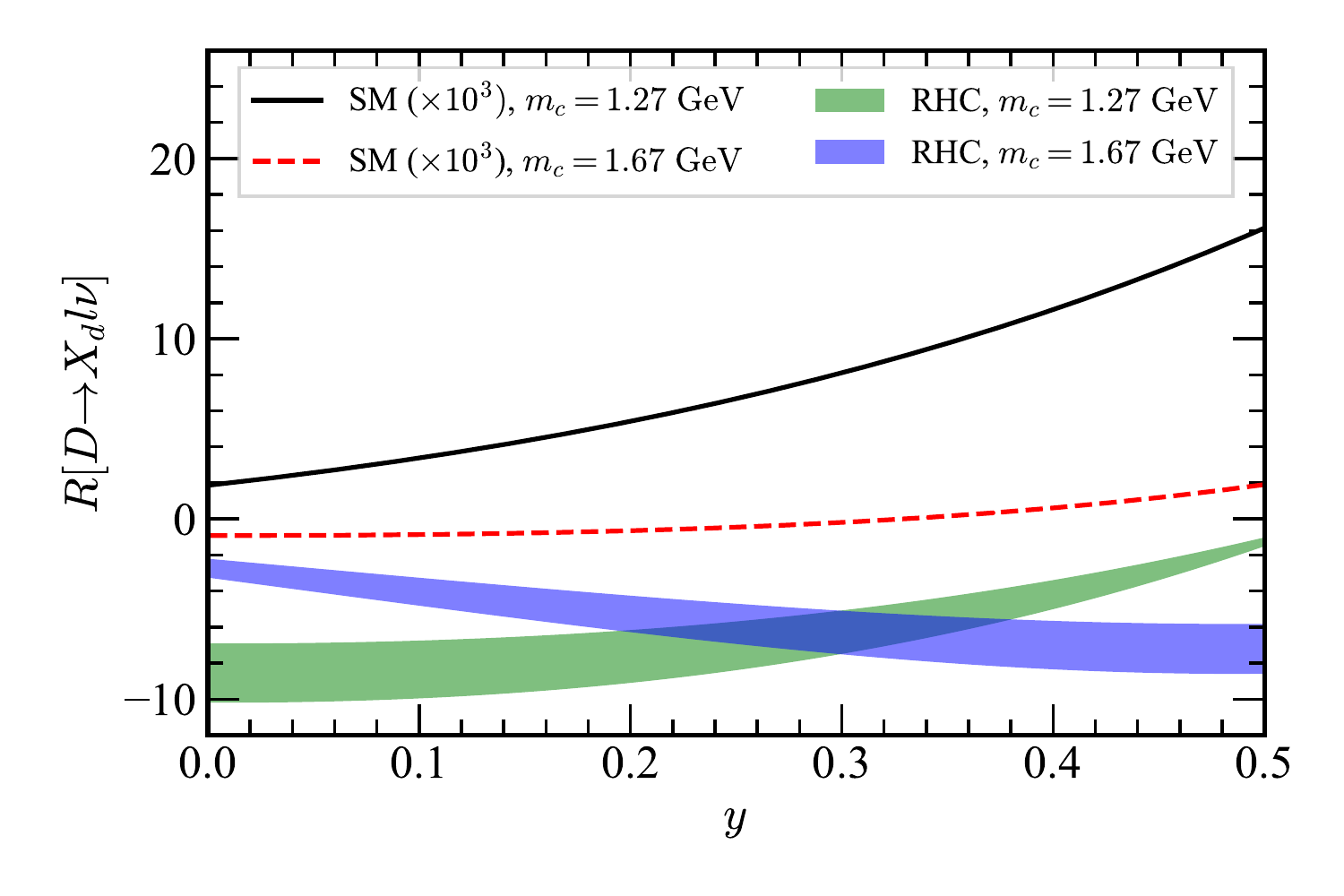}
\vspace{-10mm}
\caption{Similar to Fig.~\ref{Fig:1} except that $\bar{B}\to X_ul\nu$ is properly replaced by $D\to X_dl\nu$.}
\label{Fig:2}
\end{center}
\end{figure}

For the numerical results of the lepton energy distributions, we define a ratio similarly to the total width in Eq.~(\ref{Eq:totr}):
\bea
R_{\rm XY}[H_Q\to X_ql\nu]=\frac{\left.(\mathrm{d}\Gamma/\mathrm{d}y)\right|_{\rm XY}^{\rm I}}{\left.(\mathrm{d}\Gamma/\mathrm{d}y)\right|_{\rm XY}^{\rm pert}}.\label{Eq:refdif}
\eea
The above ratios for $(H_Q, q)= (\bar{B}, u)$ and $(D, d)$ are plotted in Figs.~\ref{Fig:1} and \ref{Fig:2}, respectively. As similarly to the total width, one can find that the duality-violating contributions for $D\to X_dl\nu$ decays are generically larger than those for $\bar{B}\to X_ul\nu$ except that the contributions vanish at oscillating nodes. It can be seen that duality violations in the SM are rather small for both $\bar{B}$ and $D$ decays. As for the RHC contributions, the typical size of the amplitude of the oscillation can be as large as the one for the perturbative contribution in $\bar{B}$ decays as shown in Fig.~\ref{Fig:1}. In the case of the $D$ decays in Fig.~\ref{Fig:2}, the duality-violating component for the RHC is even larger than that of $\bar{B}$ decays. In this regards, it should be recalled that, for the $D$ decays, the use of the saddle point approximation, including the evaluation of the external heavy quark propagation in the medium of an instanton, is less accurate than the $\bar{B}$ decays, especially for $m_c=1.27~\mathrm{GeV}$.
\vspace{-5mm}
\vspace{-10mm}
\section{Summary and discussion}
\label{Sec:IV}
In this paper, we have studied quark-hadron duality for inclusive semileptonic decays, indicating that the patterns of duality violation depend on the given effective operators. By extracting the finite-distance singularity in the light quark propagator, it is shown that the zero-mode contribution plays a pronounced role for the RHC. While the duality-violating contributions give a tiny correction to the smeared observable in the $\bar{B}$ decays, the lepton energy distribution receives a correction that is as large as the chirally suppressed perturbative contributions except for the oscillating nodes. For the $D$ decays, the typical size of duality violation is larger than the $\bar{B}$ decays, although the saddle point approximation is less accurate for these cases.
\par
The results in this work are mostly based on the single instanton inspired by the dilute nature of the instanton vacuum. Provided that the multi-instanton effects are universally treatable by the effective mass within the range estimated by the RILM and the IILM, the method is able to guarantee entire consistency at the short-distance region, where the finite-distance singularity occurs $(\sqrt{|x^2|}=2\rho)$. Obviously, in order to investigate duality violation in a more quantitative way, the very detail of the QCD vacuum, including fully interacting aspects of the instanton liquid, is to be taken into account.\par
Furthermore, a genuine instanton-induced correction to the RHC, which is not totally calculated in this work, would also play a somewhat pronounced role: This contribution is, in general, chirally allowed due to the fermionic zero mode, as well as duality violation analyzed in this work. Another point to be mentioned is that duality violation is parametrically enhanced by small $\rho$, as can be seen in Eqs.~(\ref{Eq:RHCres}), (\ref{Eq:ShifLL}), (\ref{Eq:upperone}), and (\ref{Eq:lepSM}), although the contributions of the instanton smaller than $\rho=1/3~\mathrm{fm}$ are suppressed by its small density. In order to investigate how significant the mentioned types of the contributions are, the numerical simulation based on, {\it e.g.}, the IILM should be performed.
\par
For avoiding the huge background of $\bar{B}\to X_c l\nu$ decays, the experimental data for $\bar{B}\to X_u l\nu$ decays are considered in the particular phase space region characterized by the multiscales, where a proper formalism is to be based on the soft-collinear effective theory (SCET) \cite{Bauer:2000ew}. In the analysis of this region, the instanton-induced effects should be included in the context of the SCET in a way discussed for the effective field theory \cite{Georgi:1993xi}, which gives a practical end point spectrum on $\bar{B}\to X_u l\nu$ decays.
\par
Another issue to be noted is that a problem which is originally pointed out in Ref.~\cite{Chibisov:1996wf} still remains unsolved: The method is unable to be applied for spectator-dependent diagrams such as weak annihilation or Pauli interference (the latter is for nonleptonic decays), since the propagation of external light quarks is nontrivial even at the saddle point. For improving this, the quark equation of motion in the medium of an instanton should be solved properly. Meanwhile, the generically discussed aspect for finite-distance singularity that originates from the zero mode can be also applied to effective operators such as $(\bar{l}\gamma^\mu P_{\rm L}\nu)\partial^\nu (\bar{u}\sigma_{\mu\nu} P_{\mathrm{L}}b)$ discussed in Ref.~\cite{Dassinger:2007pj} for $B\to X_cl\nu$ decays. In those specific cases, zero-mode-induced corrections to the OPE cannot be straightforwardly omitted due to their chirality structures.
\par
\begin{acknowledgments}
This work was supported in part by MOST of the Republic of China under Grant No.~MOST-110-2811-M-001-540-MY3.
\end{acknowledgments}

\end{document}